\documentclass[3p,times]{elsarticle}

\usepackage{ecrc}

\volume{00}

\firstpage{1}

\journalname{Procedia Computer Science}

\runauth{}

\jid{procs}

\jnltitlelogo{Procedia Computer Science}

\CopyrightLine{2014}{Manuscript in  Elsevier Ltd format.}

\usepackage{amssymb}

\usepackage[figuresright]{rotating}

\usepackage{epsfig}
\usepackage{epstopdf}
\usepackage[T1]{fontenc}
\usepackage{amssymb}
\usepackage{amsmath}
\usepackage{amsfonts}
\usepackage{verbatim}
\usepackage{graphicx}
\usepackage{hyperref}
\usepackage{array}

\usepackage{placeins}

\usepackage{tikz}
\usetikzlibrary{positioning}
  \usepackage[ruled]{algorithm}
\usepackage{algpseudocode}
\usepackage{lipsum}
\usepackage{multicol}
\usepackage{float}

\begin{document}

\begin{frontmatter}



\dochead{}

\title{Multi Core SSL/TLS Security Processor Architecture  Prototype Design with automated Preferential Algorithm in FPGA}


\author{Rourab Paul$^1$ 
Amlan Chakrabarti$^2$ and Ranjan Ghosh$^3$
}

\address{A.K.Choudhury School of Information Technology1,5 , Dept. of Computer Science and Engineering2 and 
Institute of Radio Physics and Electronics3,4,6, 
University of Calcutta, 92 A. P. C. Road, Kolkata–700 009, India.
}

\begin{abstract}
In this paper a pipelined architecture of a high speed network security processor (NSP) for SSL/TLS protocol is implemented on a system on chip (SOC) where hardware information of all encryption, hashing and key exchange algorithms are stored in flash memory in terms of bit files, in contrary to related works where all are actually implemented in hardware. The NSP finds applications in e-commerce, virtual private network (VPN) and in other fields that require data confidentiality. The motivation of the present work is to dynamically execute applications with stipulated throughput within budgeted hardware resource and power. A preferential algorithm choosing an appropriate cipher suite is proposed, which is based on Efficient System Index (ESI) budget comprising of power, throughput and resource given by the user. The bit files of the chosen security algorithms are downloaded from the flash memory to the partial region of field programmable gate array (FPGA). The proposed SOC controls data communication between an application running in a system through a PCI and the Ethernet interface of a network. Partial configuration feature is used in ISE14.4 suite with ZYNQ 7z020-clg484 FPGA platform. The performances of the implemented crypto algorithms are considerably better than the existing works reported in literatures. 
\end{abstract}


\end{frontmatter}


\section{Inroduction}
Security protocols, SSL, TLS and IPSec are executed in software providing run time flexibility, but executing in large time. For better throughput, many hardware architectures are proposed in literatures \cite{ieee:six}, \cite{motorola:ssl}, \cite{broadcom:ssl},  \cite{hifn:ssl} and \cite{ssl:gbps}, where the flexibility of crypto algorithms is achieved by making all algorithms available in FPGA-ASIC platform during its entire runtime. The combination of algorithms, selected for a particular session gets activated by an enable signal deactivating others allowing the deactivated ones to continue consuming resources and power. Reducing the power-resource metrics and simultaneously preserving the algorithmic flexibility is a serious challenge in embedded systems. In this paper, keeping the challenging issues in mind, it is proposed to store bit files of all necessary algorithms (encryption, hashing and key exchange algorithms) in the flash memory, instead of placing them in the reconfigurable cells (distributed RAM) of FPGA \cite{xilinx:fpga}. The bit files of the crypto suite chosen following a preferential algorithm, proposed in the paper, are only configured in the reconfigurable cells of the partitioned partial region of the FPGA saving a huge hardware slices and system power consumption. The contributions in this paper are as follows: (I) proposition of an NSP architecture using two ARM cortex processors, two crypto engines and 2 DMAs, (II) communication between system and network using two interfaces, PCI and Ethernet, (III) data from PCI and Ethernet are processed in parallel by two sets of processing elements, DMAs and crypto engines, (IV) 7 encryption algorithms, 3 hash algorithms and 3 key exchange algorithms are designed and stored in the flash memory as bit files after undertaking a thorough studies of the issues of power, resource and throughput of these algorithms, (V) a preferential algorithm is proposed to choose appropriate encryption, hash and key exchange algorithms according to the budgeted ESI comprising of power, throughput and resource or any combination of them.
\par
The organization of this paper is as follows. In Section II we have described preferential algorithm, Section 
III details out the proposed system overview. Section IV and V deals with the architectural topology of the proposed design and result-implementations respectively, and finally we conclude in Section VI.

\section{Multi Metric Preferential Algorithm}
\label{sec:multi_metric}
Multi metric algorithm is a technique in proposed SSL/TLS architecture, which can suggest one or more than one cipher suite combinations according to a budgeted system metrics of the user. If any cipher suite combination can not meet the budget, system will ask the user to increase metrics  budget.
According to the application requirements of crypto embedded systems 5 modes of priority should be there.\\
$\bullet$ \textit{Power Priority Mode}: In battery powered wireless devices it is needed to maintain the power budget at its highest priority over rest of the metrics namely throughput and resource. This mode will help to choose those cipher suites, which meets the user power budget without aware of rest of the metrics.\\ 
$\bullet$ \textit{Throughput Priority Mode}: In this mode throughput budget is at highest priority over rest of the metrics i.e. power and resource. It finds application in high speed data communication system.\\
$\bullet$ \textit{Resource Priority Mode}:In this mode the resource budget is at highest priority over rest of the metrics i.e. power and throughput. For low cost embedded systems like RFID scanner it is required to reduce the resource usage as much as possible. \\
$\bullet$ \textit{Priority Mode}: This mode does not have a single metric priority like previous modes. If user has a sensitive system where all the three metrics i.e. resource, throughput and power are considered then this mode will help the user to select the appropriate cipher suite according to its hard metrics budget.\\
\subsection{Efficient System Index Evaluation(ESI)}
Here we are proposing a new factor named Efficient System Index(ESI) which is a single parameter to measure the efficiency of all possible crypto combination of SSL/TLS system. Lets see how it comes.\\
Let $P$, $T$ and $R$ are the power, throughput and resource of the crypto algorithm respectively.
Now\\
\\
$ESI \propto T,\\
ESI \propto 1/P,\\
ESI \propto 1/R,\\$
\begin{equation}
Finally~~ESI \propto T/PR~~~~For~a~given~algorithm\\
\end{equation}
Considering the unites of P,T and R the said straight forward ESI might  not work properly. The different units of measurement of Power(mili watt), Throughput(Mbps) and resource (no. of slices) could have significant effect on ESI by shadowing others. Normalization technique is a standard well know method to scale those 3 parameters from 0 to 1 value.\\
 Instead of having P, T and R directly, we would take $P/P_{max}$, $T/T_{max}$ and $R/R_{max}$ where $P_{max}$, $T_{max}$ and $R_{max}$ represent maximum value of Power, Throughput and Resource respectively. We need to inverse the scaling range of P and R, as they are inversely proportional to ESI. So finally the ESI look like this\\
\begin{equation}
ESI=(1-\frac{P}{P_{max}}) + \frac{T}{T_{max}} +  (1-\frac{R}{R_{max}}).\\
\label{norm:equ}
\end{equation}
Now before evaluating ESI of all possible combinations of SSL/TLS we need to generate the power, throughput, and resource table from those same combinations.
 Let $E[3\times n]$, $H[3\times m]$ and $K[3\times l]$ are 3 matrices representing performance metrics of encryption, hash and key exchange algorithms respectively. The $ 1^{st}, 2^{nd}$ and $3^{rd}$ columns of these matrices represent power (mw), throughput (Mbps), and resources (no. of slices) respectively. 3 elements of each row in matrices $E$, $H$ and $K$ represent power, throughput and resource of encryption, hash and key exchange algorithm respectively. e.g. The elements of $3_{rd}$ row of $E$ matrix, $P_{3E}$, $T_{3E}$ \& $R_{3E}$ represent the power, throughput nd resource respectively of the $3_{rd} encryption algorithm of encryption list.$ of  Matrices $E$, $H$ and $K$ have $n$, $m$ and $l$ number of rows respectively, as we have $n$ number of encryption algorithms, $m$ number of hash algorithms and $l$ number of key exchange algorithms. In this work we have chosen $n$=7, $m$=3 and $l$=3. As per the SSL and TLS cipher suite referred in~\cite{Forouzan:book}, all encryption, hash and key exchange algorithms are disjoint, so all combinations of algorithms represented in $P$, $T$ and $R$ matrices are practically possible.

 \[ E=
  \left[ {\begin{array}{ccc}
P_{1E} & T_{1E} & R_{1E} \\
P_{2E} & T_{2E} & R_{2E} \\
P_{3E} & T_{3E} & R_{3E}\\
... & ... & ... \\
   P_{nE} & T_{nE} & R_{nE} \\
  \end{array} } \right]
\]
\vspace{-2 pt}%
 \[ H=
  \left[ {\begin{array}{ccc}
P_{1H} & T_{1H} & R_{1H} \\
P_{2H} & T_{2H} & R_{2H} \\
P_{3H} & T_{3H} & R_{3H}\\
... & ... & ... \\
P_{mH} & T_{mH} & R_{mE} \\
  \end{array} } \right]
\]
\vspace{-2 pt}
 \[ K=
  \left[ {\begin{array}{ccc}
P_{1K} & T_{1K} & R_{1K} \\
P_{2K} & T_{2K} & R_{2K} \\
P_{0K} & T_{0K} & R_{0K}\\
... & ... & ... \\
P_{lK} & T_{lK} & R_{lK} \\
  \end{array} } \right]
\]
Inside the iterative loops in line numbers 5, 6 and 7 of the algorithm~\ref{Algorithm:ESI} matrices $P[n\times m\times l]$, $T[n\times m\times l]$ and $R[n\times m\times l]$ have been generated to store power, throughput and resource of every combination of encryption, hash and key exchange algorithms. The logical index for each cell of the three matrices generates memory references for the corresponding encryption, hash and key exchange algorithms, e.g. $P_{012}$ refers the additive power of~$0^{th}$ encryption, $1^{st}$ hash and $2^{nd}$key exchange algorithm respectively. Figure 1 is the pictorial view of the Power matrix. Rest of the two 3d matrices i.e. $T[n\times m\times l]$ and $R[n\times m\times l]$ will look identical as Figure 1.

\begin{figure}[!htb]
\centering
\begin{tikzpicture}[every node/.style={minimum size=1cm},on grid]
\begin{scope}[every node/.append style={yslant=-0.5},yslant=-0.5]

\shade[right color=white!10, left color=white!50] (-1,-1) rectangle +(3,3);
\node at (-0.5,2.5) {$P_{111}$};
\node at (0.5,2.5) {$P_{112}$};
\node at (1.5,2.5) {$P_{11..}$};
\node at (2.5,2.5) {$P_{11l}$};

\node at (-0.5,1.5) {$P_{121}$};
\node at (0.5,1.5) {$P_{122}$};
\node at (1.5,1.5) {$P_{12..}$};
\node at (2.5,1.5) {$P_{12l}$};

\node at (-0.5,0.5) {$P_{1..1}$};
  \node at (0.5,0.5) {$P_{1..2}$};
  \node at (1.5,0.5) {$P_{1.. ..}$};
  \node at (2.5,0.5) {$P_{1..l}$};

  \node at (-0.5,-0.5) {$P_{1m1}$};
  \node at (0.5,-0.5) {$P_{1m2}$};
\node at (1.5,-0.5) {$P_{1m..}$};
  \node at (2.5,-0.5) {$P_{1ml}$};

 \draw (-1,-1) grid (3,3);
\end{scope}
\begin{scope}[every node/.append style={yslant=0.5},yslant=0.5]
  \shade[right color=white!70,left color=white!10] (3,-3) rectangle +(3,3);
  \node at (3.5,-0.5) {$P_{11l}$};
 \node at (3.5,-1.5) {$P_{12l}$};
  \node at (3.5,-2.5) {$P_{1..l}$};
 \node at (3.5,-3.5) {$P_{1ml}$};

  \node at (4.5,-0.5) {$P_{21l}$};
 \node at (4.5,-1.5) {$P_{22l}$};
  \node at (4.5,-2.5) {$P_{2..l}$};
\node at (4.5,-3.5) {$P_{2ml}$};

  \node at (5.5,-0.5) {$P_{..1l}$};
  \node at (5.5,-1.5) {$P_{..2l}$};
  \node at (5.5,-2.5) {$P_{....l}$};
 \node at (5.5,-3.5) {$P_{..ml}$};

\node at (6.5,-0.5) {$P_{n1l}$};
 \node at (6.5,-1.5) {$P_{n2l}$};
\node at (6.5,-2.5) {$P_{n..l}$};
\node at (6.5,-3.5) {$P_{nml}$};
 
\draw (3,-4) grid (7,0);
\end{scope}
\begin{scope}[every node/.append style={
    yslant=0.5,xslant=-1},yslant=0.5,xslant=-1
  ]
  \shade[bottom color=white!10, top color=white!80] (6,3) rectangle +(-3,-3);
  
\node at (3.5,3.5) {$P_{111}$};
\node at (3.5,2.5) {$P_{112}$};
  \node at (3.5,1.5) {$P_{11..}$};
  \node at (3.5,0.5) {$P_{11l}$};

\node at (4.5,3.5) {$P_{211}$};
  \node at (4.5,2.5) {$P_{212}$};
  \node at (4.5,1.5) {$P_{21..}$};
  \node at (4.5,0.5) {$P_{21l}$};

\node at (5.5,3.5) {$P_{..11}$};
  \node at (5.5,2.5) {$P_{..12}$};
  \node at (5.5,1.5) {$P_{..1..}$};
  \node at (5.5,0.5) {$P_{..1l}$};

\node at (6.5,3.5) {$P_{n11}$};
  \node at (6.5,2.5) {$P_{n12}$};
  \node at (6.5,1.5) {$P_{n1..}$};
  \node at (6.5,0.5) {$P_{n1l}$};
  \draw (3,0) grid (7,4);

\end{scope}
\end{tikzpicture}
\label{fig:power_mat}
\caption{Power Matrix}
\end{figure}
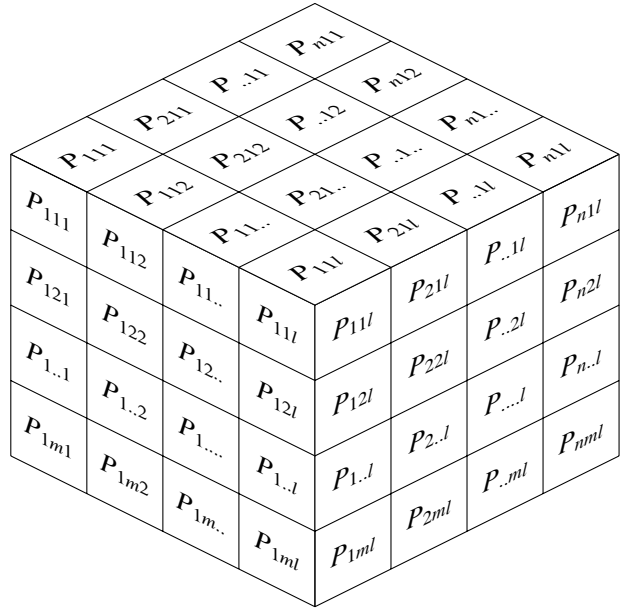
According to the requirement of user the priority of power, throughput and resource should be flexible. To put more priority to any of these three factor, we introduce 3 weighted value $W_p, W_t$ and $W_r$ to define the priority of power, throughput and resource respectively. Higher priority has been defined as more weighted value. After imposing the priority factor equation \ref{norm:equ} is altered as  
\begin{equation}
ESI=W_p\times(1-\frac{P}{P_{max}}) + W_t\times \frac{T}{T_{max}} + W_r\times (1-\frac{R}{R_{max}}).
\label{weighted:equ}
\end{equation}
As the formula are normalized $W_p+W_t+W_r=1$. Using equation \ref{weighted:equ} the Line 11 of algorithm~\ref{Algorithm:ESI} is evaluating the cut off ESI ($ESI_t$). It looks like\\
\begin{equation}
ESI_t=W_p(1-\frac{P_{avg}}{P_{max}})+W_t\frac{T_{avg}}{T_{max}}+W_r(1-\frac{R_{avg}}{R_{max}})
\label{esit:equ}
\end{equation}
where the $ P_{avg}=\frac{1}{n*m*l}\sum_{k=1}^{k=l}\sum_{j=1}^{j=m}\sum_{i=1}^{i=n}P_{ijk}$\\
$ T_{avg}=\frac{1}{n*m*l}\sum_{k=1}^{k=l}\sum_{j=1}^{j=m}\sum_{i=1}^{i=n}T_{ijk}$\\
$ R_{avg}=\frac{1}{n*m*l}\sum_{k=1}^{k=l}\sum_{j=1}^{j=m}\sum_{i=1}^{i=n}R_{ijk}$\\

In the $2^{nd}$ set of loops starting from line no. 12, 13 and 14 ESI of each set of cipher suite combinations are being calculated by equation \ref{weighted:equ}, and after that at line no 15 calculated ESI has been compared to previous $ESI_t$ to suggest eligible cipher suite combinations.
Considering the priority of power, throughput and resource ($W_p, W_t$ and $W_r$) 3 category has been set. Only one decimal digit at left side of point has been considered as more than one decimal digit after point could not effect the ESI parameter significantly. The priority categories of priority classification
\begin{enumerate} 
\item \textit{equal priority}: where $W_p, W_t$ and $W_r$ has same weight.$W_p$=$W_t$=$W_r$.
\item \textit{single priority}: where system cares only about single parameter priority over the rest of the parameters. e.g. if we have only power priority then $W_p$=1, $W_t$=0 and $W_r$=0. This is relevant to section\ref{sec:multi_metric} Power, Resource and throughput Priority mode.
\item \textit{multiple priority}: when system is concerned about more than one parameter. For such cases all weight $W_p\neq0, W_t\neq0$ and $W_r\neq0$.  This is relevant to section\ref{sec:multi_metric} Priority mode.
\end{enumerate}
Table \ref{table:weight_table} shows the 3 categories of priority where $1 ^{st}$ instance is for equal priority, $2^{nd}$ to $4^{th}$ are for single priority and $5^{th}$ to $46^{th}$ are for multiple priority category. For  $1^{st}$ instance, we have equal priority on power, throughput and resource where best and worst algorithm combinations are  DES+MD5+RSA and AES+SHA256+DH\_RSA respectively depending on its ESI values. For $2^{nd}$, $3^{rd}$ and $4^{th}$ instances the priorities are on power, throughput and resource respectively. As we have only priority on power in $2^{nd}$ instance, only $W_p$ is '1' and rest of the weight $W_t$ and $W_r$ are '0'. Following the same contrast $W_t$ and $W_r$ are 1 at  $3^{rd}$ and $4^{th}$ rows respectively. If you see tables \ref{table:hash_table}, \ref{table:encryption_table} and \ref{table:KeyExchg_table} of section \ref{res_impl} you can justify the best and worst algorithm combination of instances
$1^{st}$, $2^{nd}$, $3^{rd}$ and $4^{th}$ rows at table \ref{table:weight_table}. 

\makeatletter
\def\BState{\State\hskip-\ALG@thistlm}
\makeatother
\alglanguage{pseudocode}
\begin{algorithm}[]

\caption{Calculation of ESI}
\label{Algorithm:ESI}
\begin{algorithmic}[1]
\Procedure{$\mathbf{Input Matrix}$}{$E[3\times n]; H[3\times m]; K[3\times l]$}
    \For {$i = 1 \to n$}
             \For {$j = 1 \to m$}
                 \For {$k = 1 \to l$}.
                     \State P[i,j,k]=E[i,0]+H[j,0]+K[k,0];
                     \State T[i,j,k]=E[i,1]+H[j,1]+K[k,1];
                     \State R[i,j,k]=E[i,2]+H[j,2]+K[k,2];
                \EndFor
            \EndFor
    \EndFor
\State$ESI_t=W_p(1-\frac{P_{avg}}{P_{max}})+W_t\frac{T_{avg}}{T_{max}}+W_r(1-\frac{R_{avg}}{R_{max}})$
\if
\State jgjgj
\fi
    \For {$i = 1 \to n$}
             \For {$j = 1 \to m$}
                 \For {$k = 1 \to l$}
 \State ESI[i,j,k]=$W_p(1-\frac{P[i,j,k]}{P_{max}})+W_t\frac{T[i,j,k]}{T_{max}}+W_r(1-\frac{R[i,j,k]}{R_{max}})$
\If {$ESI_t=<ESI[i,j,k]$} 
\State $ \backslash\backslash\textit{Eligible Combinations are:}$
\State $---------------$
\State $Encryption\_Algorithm[i]$
\State $Hash\_Algorithm[j]$
\State $Key\_Exchange\_Algorithm[k]$
\State $---------------$
\EndIf
                  \EndFor
            \EndFor
    \EndFor

\EndProcedure

\Statex

\end{algorithmic}

\end{algorithm}


%
%
%
%
%
\begin{figure}[!hp]
\centering
\vspace{-9pt}
\includegraphics[width=9cm, height=8cm]{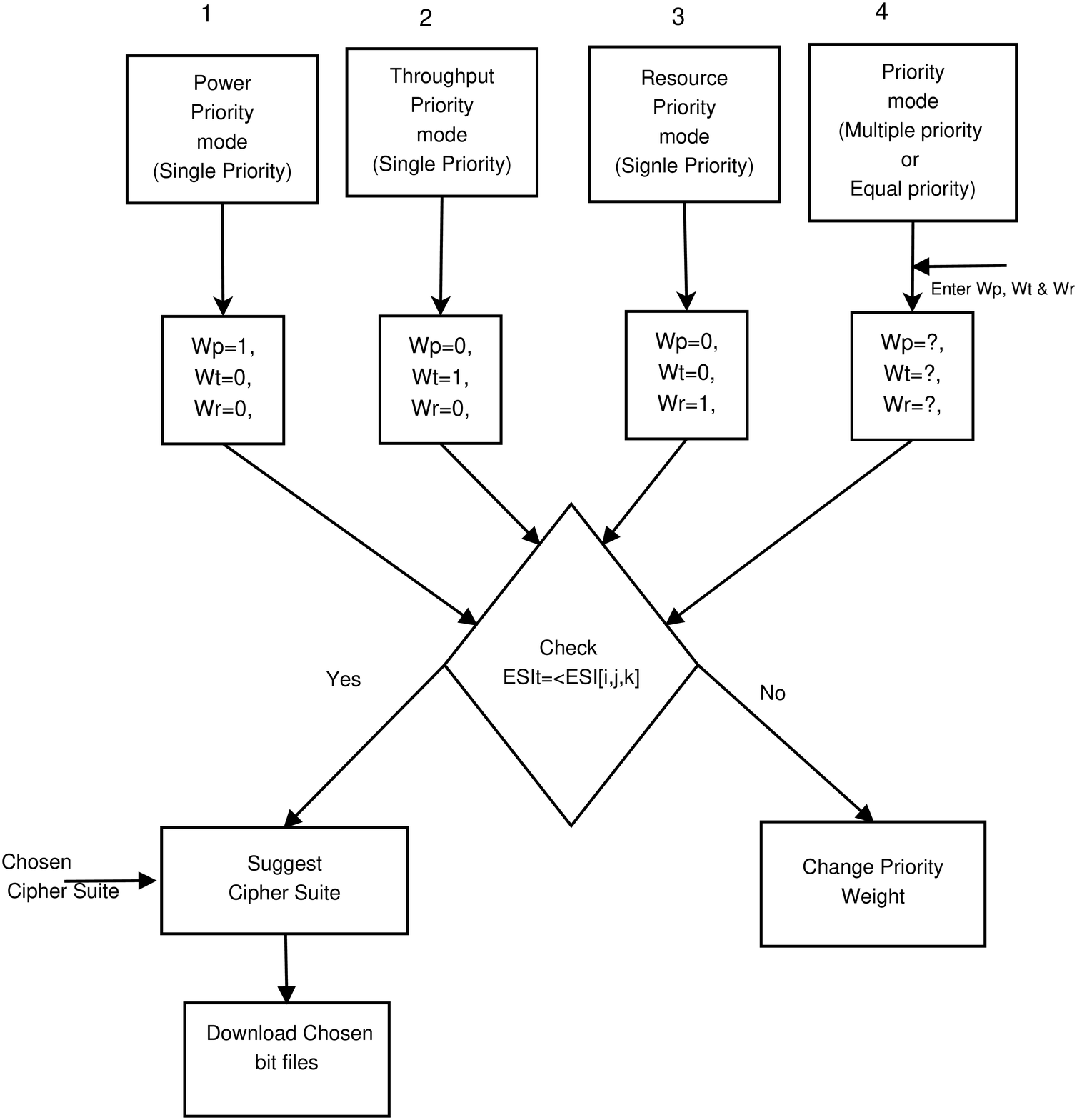}
\vspace{-6pt}
\caption{Modes of Preferential Algorithm}
\label{fig:mode_fig}
\end{figure}
 
\begin{figure}[!hp]
\centering
\vspace{1pt}
\includegraphics[width=9cm, height=6cm]{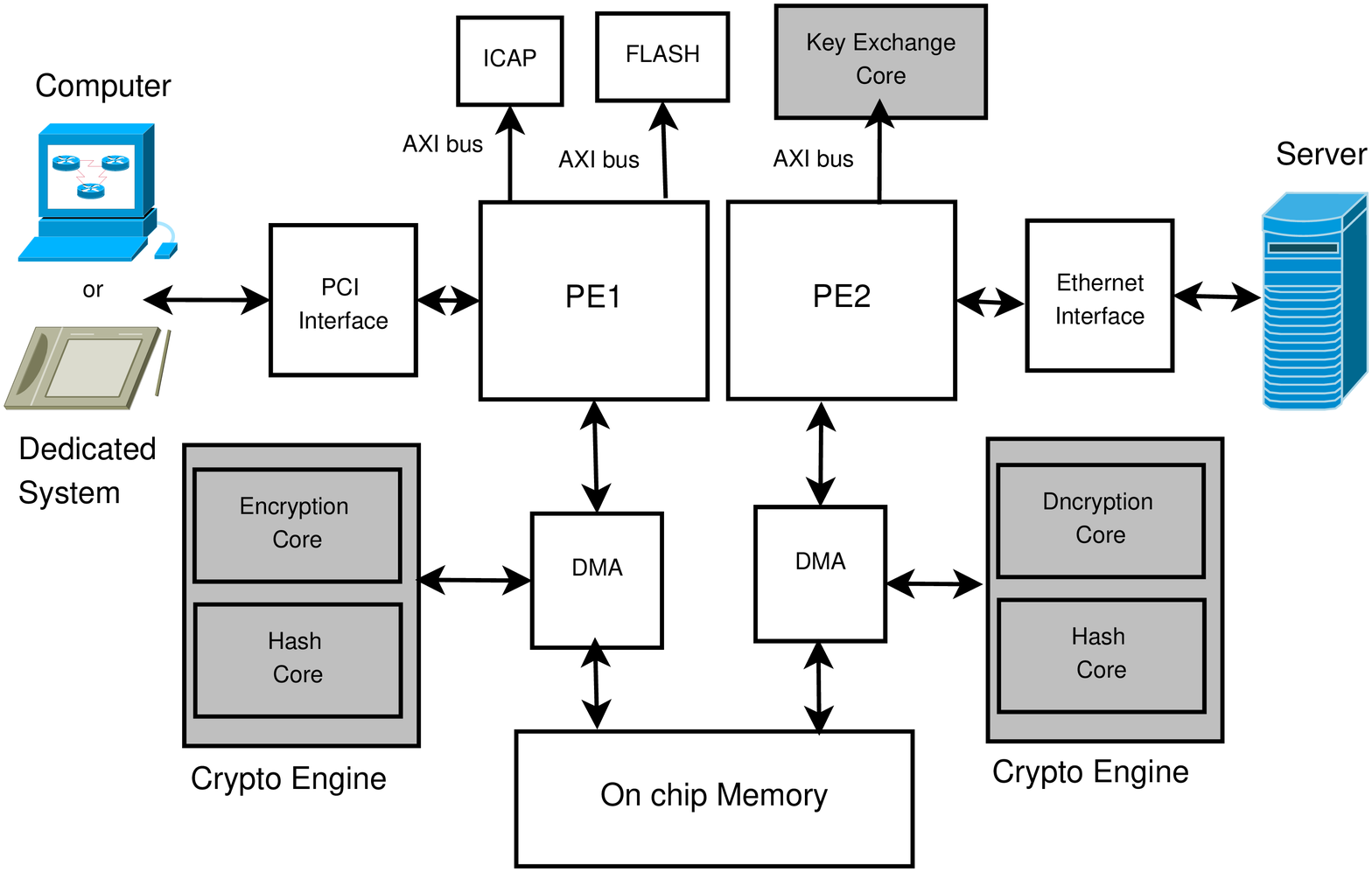}
\vspace{-6pt}
\caption{System Architecture}
\vspace{8pt}
\label{fig:system_arch_fig}
\end{figure}

\begin{table*}[!hp]
\caption{Preferential algorithm} 
\centering  
\resizebox{17cm}{!}{%

    \begin{tabular}{|c|c|c|c|c|c|c|c|c| }
        \hline
Sl. & \multicolumn{3}{|c|}{Weight}   &Priority&~      &\multicolumn{2}{c|}{Combination}&\% of Eligible\\\cline{2-4}\cline{7-8}
~&$W_p$       & $ W_t$ &	       $W_r$&~&	       $ESI_t$   &    Best& Worst&Combination\\\hline
\hline
1&0.333&       0.333&	          0.333&equal&		0.3398&     DES+MD5+RSA&AES+SHA256+DH\_RSA&46\\\hline
\hline
2&1&	       0&	         0&	Single&	          0.3098&     Idea+MD5+RSA&AES+SHA512+DH\_anon&71.4\\
\hline
3&0&             1&	          0&	Single&	          0.3713&    DES+SHA512+RSA&Idea+SHA256+DH\_RSA&38.1\\
\hline
4&0&             0&	          1&	Single&	          0.3384&    Grain+MD5+RSA&AES+SHA512+DH\_RSA&66.6\\\hline
\hline
5&0.8&         0.1&	        0.1&	multiple&		0.3188&    DES+MD5+RSA&AES+SHA512+DH\_RSA&68.2\\\hline
6&0.7&         0.2&	        0.1&	multiple&		0.325&    DES+MD5+RSA&AES+SHA512+DH\_RSA&65\\\hline
7&0.7&         0.1&	        0.2&	multiple&		0.3217&    DES+MD5+RSA&AES+SHA512+DH\_RSA&68.2\\\hline
8&0.7&         0.15&	        0.15&	multiple&		0.3233&    DES+MD5+RSA&AES+SHA512+DH\_RSA&68.2\\\hline
9&0.6&         0.2&	        0.2&	multiple&		0.3278&    DES+MD5+RSA&AES+SHA512+DH\_RSA&65\\\hline
10&0.6&         0.3&	        0.1&	multiple&		0.3311&    DES+MD5+RSA&AES+SHA256+DH\_RSA&52.3\\\hline
11&0.6&         0.1&	        0.3&	multiple&		0.3245&    DES+MD5+RSA&AES+SHA512+DH\_RSA&71.4\\\hline
12&0.5&         0.3&	        0.2&	multiple&		0.3339&    DES+MD5+RSA&AES+SHA256+DH\_RSA&55.5\\\hline
13&0.5&         0.2&	        0.3&	multiple&		0.3306&    DES+MD5+RSA&AES+SHA512+DH\_RSA&65\\\hline
14&0.5&         0.25&	        0.25&	multiple&		0.3323&    DES+MD5+RSA&AES+SHA512+DH\_RSA&58.7\\\hline
15&0.5&         0.4&	        0.1&	multiple&		0.3373&    DES+MD5+RSA&AES+SHA256+DH\_RSA&42.8\\\hline
16&0.5&         0.1&	        0.4&	multiple&		0.3274&    DES+MD5+RSA&AES+SHA512+DH\_RSA&71.4\\\hline
17&0.4&         0.3&	        0.3&	multiple&		0.3368&    DES+MD5+RSA&AES+SHA512+DH\_RSA&55.5\\\hline
18&0.1&             0.8&	          0.1&	multiple&	          0.3618&    DES+SHA512+RSA&Idea+SHA256+DH\_RSA&41.2\\\hline
19&0.2&             0.7&	          0.1&	multiple&	          0.3557&    DES+SHA512+RSA&AES+SHA256+DH\_RSA&42.8\\\hline
20&0.1&             0.7&	          0.2&	multiple&	          0.3586&    DES+SHA512+RSA&AES+SHA256+DH\_RSA&42.8\\\hline
21&0.15&             0.7&	          0.15&	multiple&	          0.3571&    DES+SHA512+RSA&AES+SHA256+DH\_RSA&42.8\\\hline
22&0.2&             0.6&	          0.2&	multiple&	          0.3524&    DES+SHA512+RSA&AES+SHA256+DH\_RSA&42.8\\\hline
23&0.3&             0.6&	          0.1&	multiple&	          0.3496&   DES+SHA512+RSA&AES+SHA256+DH\_RSA&42.8\\\hline
24&0.1&             0.6&	          0.3&	multiple&	          0.3553&   DES+SHA256+RSA&AES+SHA256+DH\_RSA&42.8\\\hline
25&0.3&             0.5&	          0.2&	multiple&	          0.3462&    DES+SHA512+RSA&AES+SHA256+DH\_RSA&42.8\\\hline
26&0.2&             0.5&	          0.3&	multiple&	          0.3491&    DES+SHA512+RSA&AES+SHA256+DH\_RSA&42.8\\\hline
27&0.4&             0.5&	          0.1&	multiple&	          0.3434&   DES+SHA512+RSA&AES+SHA256+DH\_RSA&42.8\\\hline
28&0.1&             0.5&	          0.4&	multiple&	          0.352&     DES+SHA512+RSA&AES+SHA256+DH\_RSA&42.8\\\hline
29&0.25&             0.5&	          0.25&	multiple&	          0.3477&    DES+SHA512+RSA&AES+SHA256+DH\_RSA&42.8\\\hline
30&0.3&             0.4&	          0.3&	multiple&	          0.3430&    DES+SHA512+RSA&AES+SHA256+DH\_RSA&42.8\\\hline
31&0.1&             0.1&	          0.8&	multiple&	          0.3388&    Grain+MD5+RSA&AES+SHA512+DH\_RSA&69.8\\\hline
32&0.2&             0.1&	          0.7&	multiple&	          0.3359&    DES+MD5+RSA&AES+SHA512+DH\_RSA&69.8\\\hline
33&0.1&             0.2&	          0.7&	multiple&	          0.3421&    DES+MD5+RSA&AES+SHA512+DH\_RSA&68.2\\\hline
34&0.15&            0.15&         0.7&	multiple&	          0.339&    DES+MD5+RSA&AES+SHA512+DH\_RSA&71.4\\\hline
35&0.2&             0.2&	          0.6&	multiple&	          0.3392&    DES+MD5+RSA&AES+SHA512+DH\_RSA&68.2\\\hline
36&0.3&             0.1&	          0.6&	multiple&	          0.3331&   DES+MD5+RSA&AES+SHA512+DH\_RSA&69.8\\\hline
37&0.1&             0.3&	          0.6&	multiple&	          0.3454&    DES+MD5+RSA&AES+SHA512+DH\_RSA&53.9\\\hline
38&0.3&             0.2&	          0.5&	multiple&	          0.3364&    DES+MD5+RSA&AES+SHA512+DH\_RSA&68.2\\\hline
39&0.4&             0.1&	          0.5&	multiple&	          0.3302&    DES+MD5+RSA&AES+SHA512+DH\_RSA&71.4\\\hline
40&0.1&             0.4&	          0.5&	multiple&	          0.3487&   DES+MD5+RSA&AES+SHA256+DH\_RSA&42.8\\\hline
41&0.2&             0.3&	          0.5&	multiple&	          0.3425&    DES+MD5+RSA&AES+SHA512+DH\_RSA&55.8\\\hline
42&0.25&           0.25&         0.5&	multiple&	          0.3394&    DES+MD5+RSA&AES+SHA512+DH\_RSA&65\\\hline
43&0.3&             0.3&	          0.4&	multiple&	          0.3397&    DES+MD5+RSA&AES+SHA512+DH\_RSA&55.5\\\hline
44&0.4&             0.2&	          0.4&	multiple&	          0.3335&   DES+MD5+RSA&AES+SHA512+DH\_RSA&65\\\hline
45&0.2&             0.4&	          0.4&	multiple&	          0.3458&    DES+MD5+RSA&AES+SHA256+DH\_RSA&42.8\\\hline
46&0.4&             0.4&	          0.2&	multiple&	          0.3401&    DES+SHA512+RSA&AES+SHA256+DH\_RSA&42.8\\

\hline
    \end{tabular}
}
\label{table:weight_table} 
\end{table*}

\section{System Overview}
The proposed hardware architecture of SSL/TLS protocol is a prototype single chip solution for parallel, multi way data communication, which may be a replacement of SSL software application in our general purpose computer and low budget embedded gadget. The proposed system would be placed between a network and embedded system/general purpose machine. Two processing elements are dedicated to receive/send data from/to special/general purpose machine and any network. The PCI and Ethernet interface is used to communicate special/general purpose machine with PE1 and network with PE2 respectively. There may be some alternative interfaces of PCI while special purpose dedicated system will be used  instead of general purpose machine. There are two data flow paths.
\begin{itemize}
\item The data coming from PCI, will be written in the on-chip-memory using DMA, next the Crypto Engine1 (CE1) encrypts the data fetching it from the memory. The encrypted data buffer gets written on the memory again using DMA. The Processor Element 2 (PE2) reads memory through another DMA process and then sends it to the network by the ethernet interface.
\item The encrypted data coming from Ethernet, gets written in the on chip memory using DMA. That data buffer is fetched by the Crypto Engine2 (CE2) for the decryption process. The decrypted data gets written back to the memory using DMA. The Processing Element 1 (PE1) will read that data buffer using another DMA and sends it to the system using the PCI interface.
\end{itemize} 
The architectural bird view has been shown in Figure~\ref{fig:system_arch_fig}.
A brief overview of the different blocks of the proposed architecture is given below.
\subsection{Processing Element}
Two 32 bit ARM cores are used as Processing Elements (PEs) to handle the data coming from PCI, Ethernet interfaces and forwarding the data back to the memory using DMA module. PE1 selects the suitable algorithm combination using preferential algorithm to download the corresponding bit files of chosen algorithm combination into the partitioned partial region of the FPGA from the Flash memory. PE2  is responsible for key exchange procedure with server using Key Exchange Block. Both PE1 and PE2 can also synchronize the whole pipeline process using the Process  Scheduler IP (PS). Microblaze processor may be also an alternative of ARM processor as PE.
\subsection{Process  Synchronizer (PS)}
PS is a custom logic IP, which synchronizes and monitors the whole system flow and system status using its master bus and slave bus respectively. The PS is connected with AXI Streamer and Crypto Engine through controlling I/Os, which act like flag bits inside PS.
PEs, AXI Streamer and Crypto Engine can read and write those flags through master bus and slave bus of PS respectively. During a full data process cycle flags are being updated by $done$ signals of each process like $write DMA$ process, $read DMA$ process and $crypto$ process. PEs read those $done$ flags using API's of PS through AXI bus and according to the status of those $done$ signals PEs synchronize the $start$ signals of those processes. Two PS has been used by two PEs. The PS can be used for debugging purpose also to monitor the whole system status.


\begin{figure}[!hp]
\centering
\includegraphics[width=9cm, height=7.2cm]{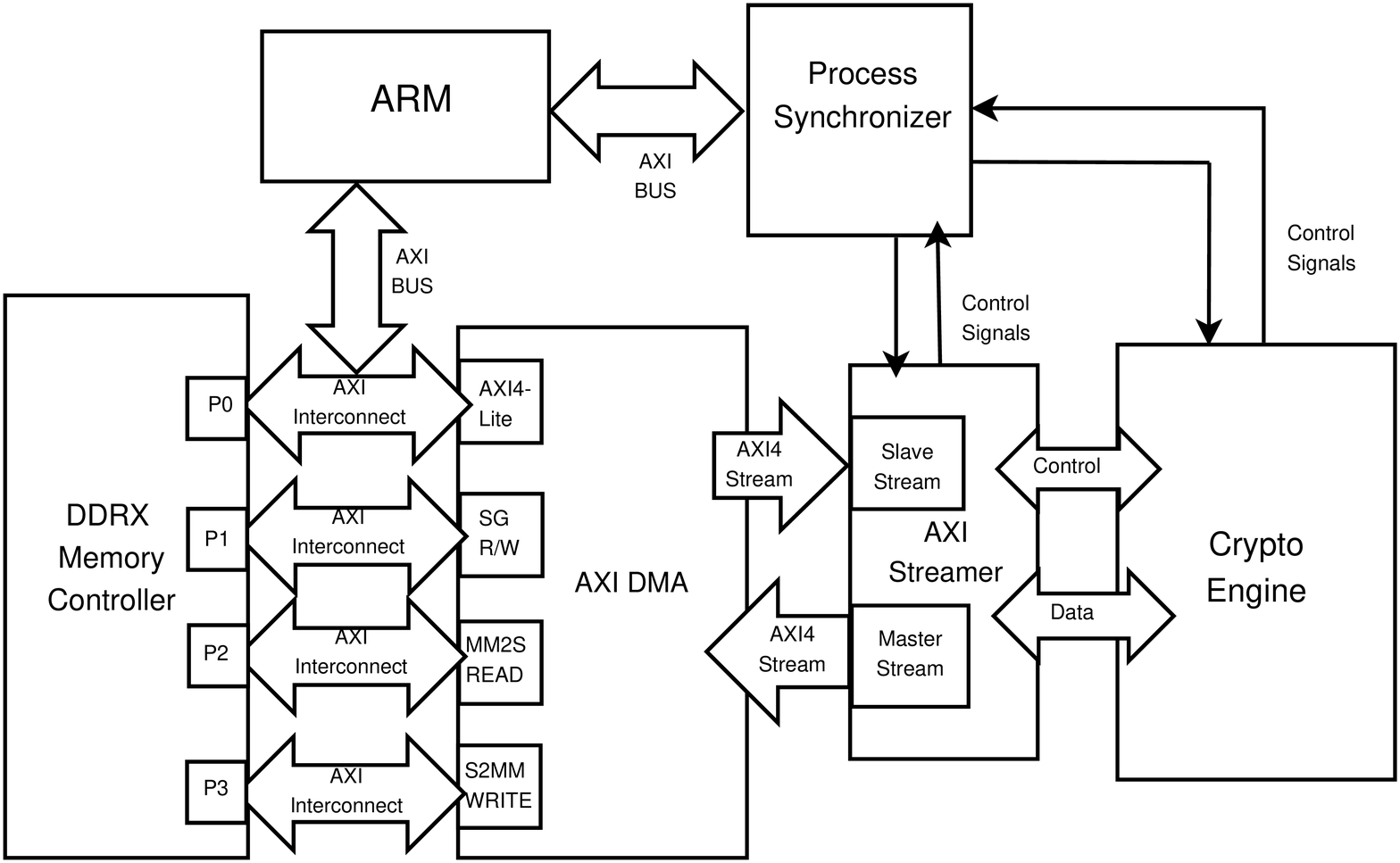}
\vspace{-6pt}
\caption{Processor and Crypto Engine Communication}
\vspace{-8pt}
\label{fig:sys_config_fig}
\end{figure}
\subsection{Crypto Engine}
Each crypto engines consists two blocks, hashing block, and encryption block. Two crypto engines have been used in two sides of the proposed architecture. One is dedicated for ciphering and hashing of plain texts coming from PCI, another is used for deciphering and hash checking of encrypted text coming from the Ethernet. 
\vspace{-10pt}
\subsection{Key Exchange Blocks}
Before starting the data communication between client and server key exchange is necessary. Since server data comes through the Ethernet interface, which is controlled by PE2, so key exchange block is also connected with PE2.
\vspace{-10pt}
\subsection{AXI Streamer}
AXI Streamer is used as a interface controller of crypto engine and DMA. DMA uses a master and slave bus for read write operation and few controlling ports to control the bus status. The details of the communication between processors and crypto engine through DMA and AXI Streamer is shown in Figure~\ref{fig:sys_config_fig}.\\  
\vspace{-10pt}
\subsection{ICAP and Flash Memory}
The Internal Configuration Access Port (ICAP) is a IP provided by xilinx to read and write bit files from any storage memory. After selection of appropriate cipher suite by preferential algorithm, the bit files of the chosen key exchange, encryption and hash algorithm gets configured from flash memory. The whole process is controlled by PE1.
Details shown in \cite{xilinx:hwicap}
The flow of the steps are show in Figure~\ref{fig:topology_our_fig}.

\section{Architecture Topology}
\begin{figure}[!hp]
\centering
\vspace{-9pt}
\includegraphics[width=9cm, height=8.2cm]{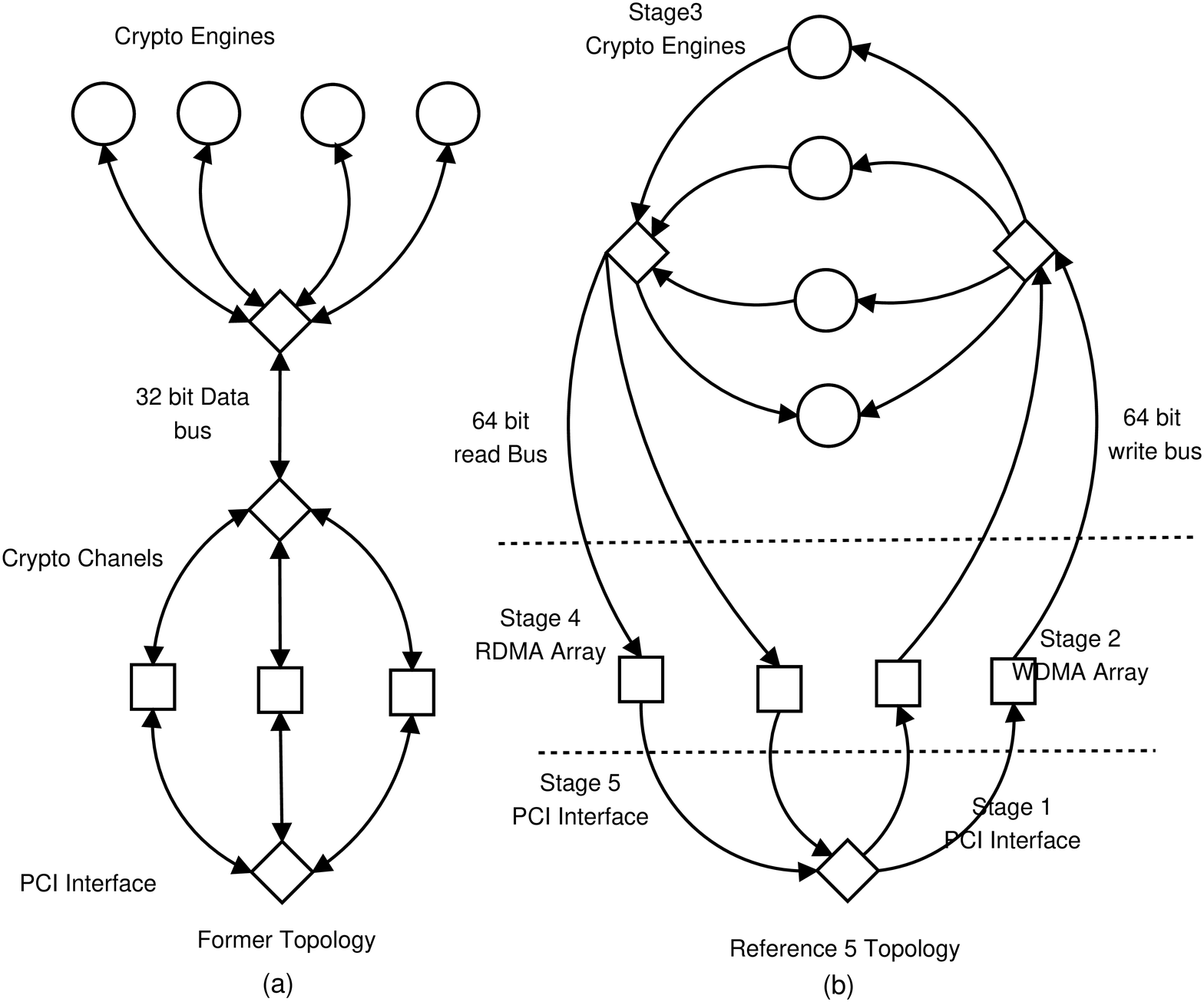}
\vspace{-6pt}
\caption{Existing Topology} 
\label{fig:topology_ur_fig}
\end{figure}

Former topology shown in Figure~\ref{fig:topology_ur_fig}(a) has been adopted by many literatures like ~\cite{ieee:six},~\cite{motorola:ssl},~\cite{broadcom:ssl} and~\cite{hifn:ssl} where a single 32 bit bidirectional bus has been used to communicate between PCI and crypto engines. Bidirectional data bus performs both data read and write operation. Even huge amount of arbitrations has been incorporated  in this kind of architecture, still the data is being congested as shown in Figure~\ref{fig:data_tran_fig}(c). The entire bus is being busy until a full data transfer has been completed. In~\cite{ssl:gbps} a pipelined architecture has been proposed where two different 64 bit bus is used for reading and writing operation, as a result both operations can be done at the same time. The authors used a single PCI interface which accepts and forwards the data after processing through the same PCI channel, which results to congestion. In Figure~\ref{fig:data_tran_fig}(b) for $5^{th}$ packet time $t$ has been compromised due the congestion of data in the PCI. Generally in most of the application systems (client) Ethernet port is the interface for transmission and reception of data. Hence, we have used of two separate interfaces as shown in Figure~\ref{fig:data_tran_fig}(a). As two separate interfaces has been used, collision of data receiving process and transmitting process has been solved in our proposed architecture, which causes an acceleration in cryptographic processes over receiving and transmitting way data path.

\subsection{PCI to Ethernet}
\begin{figure}[!hp]
\centering
\includegraphics[width=8.8cm, height=13cm]{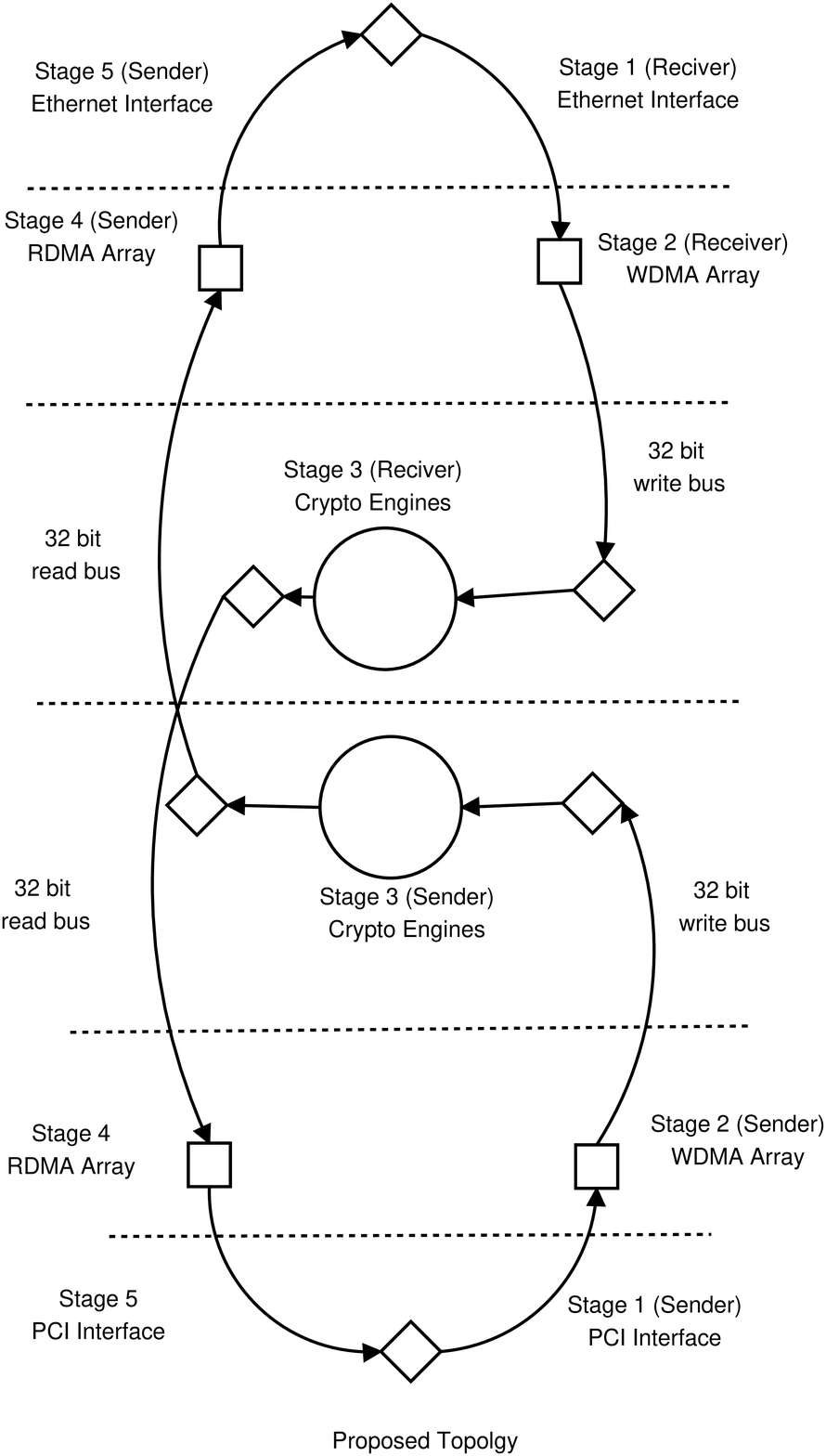}
\vspace{-6pt}
\caption{Proposed Topology}
\vspace{-8pt}
\label{fig:topology_our_fig}
\end{figure}

\begin{figure}[!hp]
\centering
\includegraphics[scale=0.30]{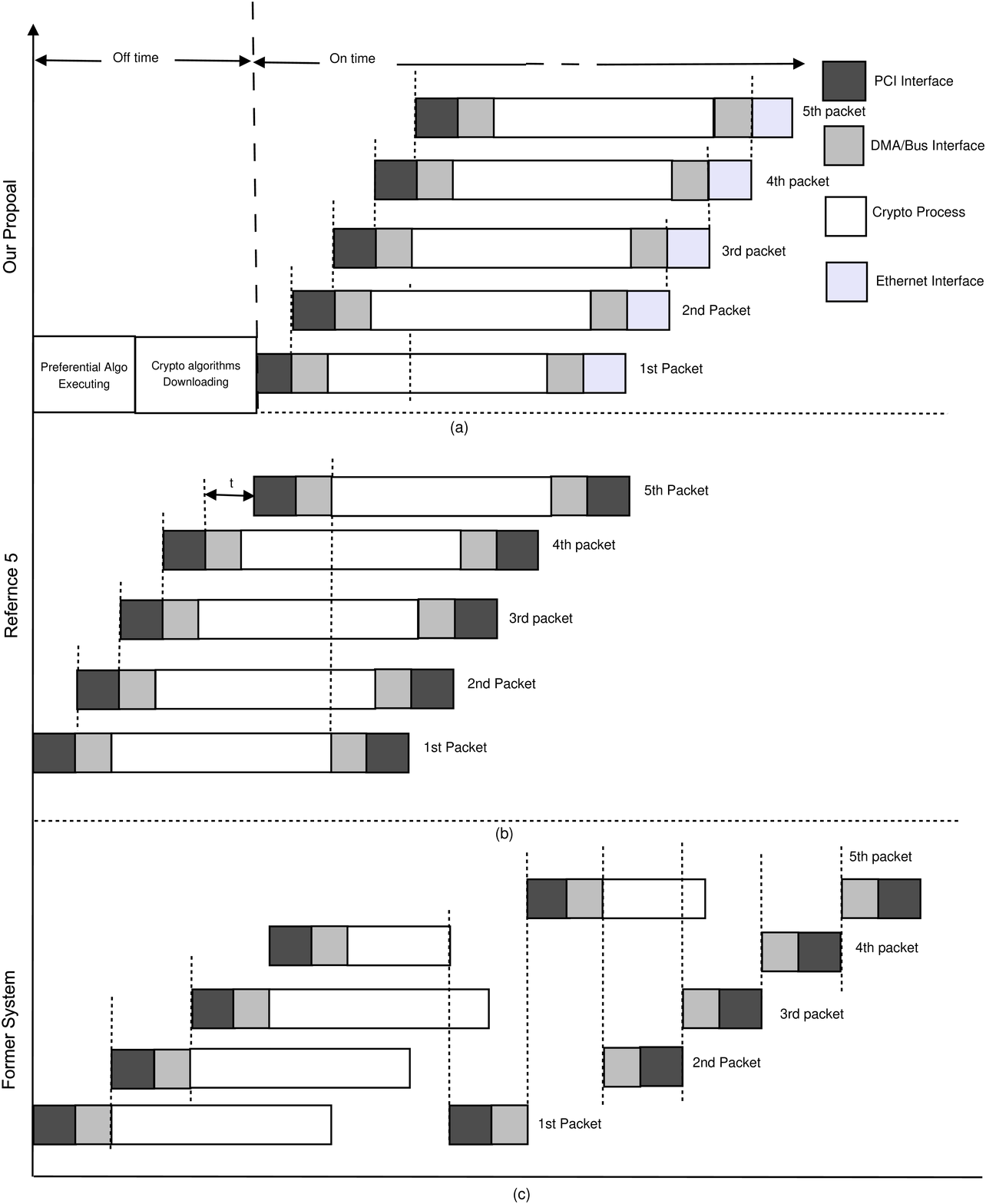}
\vspace{-6pt}
\caption{Data transfer efficiency comparison.}
\vspace{-8pt}
\label{fig:data_tran_fig}
\end{figure}

Receive plain text from system through PCI for encryption and hashing process and send it to Ethernet. From stage1 to 3 of figure \ref{fig:topology_our_fig} data are coming from PCI to Crypto engine through a Write DMA(WDMA). At stage 4, Read DMA(RDMA) read data from said crypto engine and at stage 5 send it to ethernet Interface.
\subsection{Ethernet to PCI}
Receive cipher text from ethernet for decryption and hash checking process and send it to system through PCI. Rest of the architecture topology is symmetric with previous data flow. As two separate interfaces has been used, collision of data receiving process and transmitting process has been solved in our proposed architecture

\section{Result and Implementation}
\label{res_impl}
The proposed SSL/TLS cipher suite has been implemented with 3 hashing algorithms, 7 encryption algorithms and 3 key exchange algorithms.  Different design metrics of these algorithms are shown in table~\ref{table:hash_table},~\ref{table:encryption_table} and~\ref{table:KeyExchg_table}. The possible combinations for the 3 types of algorithms are 7x3x3=63.
\begin{table}[!h]
\caption{Onchip resource throughput power of HASH algorithm} 
\centering  
\resizebox{9cm}{!}{%

    \begin{tabular}{|c|c|c|c|c|  }
        \hline
Name &	                    Slice &	           Power&	             Throughput&      Critical\\
of hash&	                   \# &	          (mw) &	              Gbps&                path(ns)\\\hline

SHA-256&	                  1385&	          		176&			0.735&	     3.85\\\hline
SHA-512&		       2647&			278&		          1.471&	         5.50\\\hline
MD5      &   	                  992&			112&			0.916&	         7.31\\

\hline
    \end{tabular}
}
\label{table:hash_table} 
\end{table}
\begin{table}[!h]
\caption{Onchip resource throughput power of Crypto algorithm} 
\centering  
\resizebox{9cm}{!}{%

    \begin{tabular}{|c|c|c|c|c|  }
        \hline
Name &	                    Slice &	           Power&	             Throughput&      Criticalt\\
of Encryption&	        \# &	          (mw) &	              Gbps&                path(ns)\\\hline

AES&	                  11385&	          		1183&			1.067&		      2.939\\\hline
RC4&		       5383&			994&		          0.931&	         4.127\\\hline
Grain &   	       237&			99.7&			0.116&	         1.689\\\hline
Salsa&		       2839&			107&		  3.725&	         2.064\\\hline
DES&		       456&			103&		          7.45&	                   2.468\\\hline
3DES&		 1478&			117&		  2.48&	         2.468\\\hline
Idea&		       320&			95&		          0.079&	         8.18\\

\hline
    \end{tabular}
}
\label{table:encryption_table} 
\end{table}
\begin{table}[!h]
\caption{Onchip resource throughput power of Key Exchange algorithm} 
\centering  
\resizebox{9cm}{!}{%

    \begin{tabular}{|c|c|c|c|c|  }
        \hline
Name &	                    Slice &	           Power&	             Throughput&      Criticalt\\
of hash&	                   \# &	           (mw) &	             Gbps&              path(ns)\\\hline
RSA&	                             13910&	          		1589&			0.298&		      3.85\\\hline
DH\_anon&		       14012&			1767&		          0.149&	         3.89\\\hline
DH\_RSA&   	                  14789&			1918&			0.099&    5.67\\

\hline
    \end{tabular}
}
\label{table:KeyExchg_table} 
\end{table}

\begin{table}[!h]
\caption{Comparison Table} 
\centering  
\resizebox{9cm}{!}{%

    \begin{tabular}{|c|c|c|c|c|c|c|  }
        \hline
Name & \cite{ssl:gbps} &	   \cite{motorola:ssl}&  \cite{ieee:six}&      \cite{ieee:seventeen}   &\cite{ieee:sixteen}  & This work\\\hline
Device&  xc3s500e&	    ASIC&		      ASIC&		      xc2vp100&                  xcv1000e                   &xc7z020\\\hline
Clock&   150&		    66&		                80&	                106.6&                               24.2                     &125\\
MHz&     ~&			   ~&		                ~&	                           ~&                                   ~                            &~\\\hline
AES &    2.042&		   --&		    	    1.462&	               1.2&                             0.310                          &1.067\\
Gbps&     /0.951&		   ~&		               ~&	                          ~&                                    ~                            &~\\\hline
DES&      2.453&		   1.13&		     --&	                          -- &                                  --                             &7.45\\
Gbps&     ~&			   ~&		               ~&	                          ~&~                                                                &~\\\hline
3DES&    0.663&		   --&		               --&	                          --&                                  --                              &2.48\\
Gbps&     ~&			   ~&		              ~&	                          ~&                                  ~                              &~\\\hline
RSA&      2350/s&	              520/s&		   26/s&	               -- &                                 --                              &0.0391gb/s\\

\hline
    \end{tabular}
}
\label{table:comp_table} 

\end{table}

%
%
%
%
\section{Conclusion}
In this paper practical application focusing the adoption of high speed pipelined NSP architecture to accelerate the performance of SSL/TLS protocol is proposed, which can be used in general purpose and also in special purpose machines. The partitioned partial region where algorithm combinations are being dumped must be made of FPGA, but rest of the system can be made of ASIC or of FPGA. The partial reconfiguration feature of FPGA can dynamically alter the security paradigms depending upon the choice of the proposed preferential algorithm regarding the requirement of power, throughput, resource and ESI of the system. The results obtained following the implementation of the proposed architecture show that it has better pipelined process, better hardware flexibility and well optimized resource-power parameters in comparison to what was achieved in any of the existing literatures.
\vspace{-8pt}

\bibliographystyle{elsarticle-num}
\bibliography{IEEEexample}

\end{document}